\documentclass[prb,notitlepage,floatfix]{revtex4-1}

\usepackage{amsfonts}
\usepackage{amsmath}
\usepackage{amssymb}
\usepackage{graphicx}
\usepackage{bm}
\usepackage{dcolumn}

\begin{document}

%\linenumbers

	\title{Gossamer bulk high-temperature superconductivity in FeSe}
	\author{A.A.~Sinchenko$^{1,2}$,P.D.~Grigoriev$^{3,4,5\ast }$, A.P.~Orlov$^{1}$,
		A.V.~Frolov$^{1}$, A.~Shakin$^{5}$, D.A.~Chareev$^{6,7}$, O.S.~Volkova$^{5,6,8}$ and A.N.~Vasiliev$^{5,6,8}$}

	\address{$^{1}$Kotel'nikov Institute of Radioengineering and Electronics of
		RAS, Mokhovaya 11-7, 125009 Moscow, Russia}
	
	\address{$^{2}$National Research Nuclear University (MEPhI), 115409
		Moscow,Russia}
	
	\address{$^{3}$L.D. Landau Institute for Theoretical Physics, 142432
		Chernogolovka, Russia}
 \email{grigorev@itp.ac.ru}
 	
    \address{$^{4}$P.N. Lebedev Physical Institute, RAS, 119991, Moscow, Russia}
	
	\address{$^{5}$National University of Science and Technology "MISiS", 119049
		Moscow, Russia}

\address{$^{6}$Ural Federal University, 620002 Ekaterinburg, Russia}
	
\address{$^{7}$Institute of Experimental Mineralogy, RAS, 142432
		Chernogolovka, Russia}

\address{$^{8}$M.V. Lomonosov Moscow State University, 119991 Moscow, Russia}	
		
\maketitle

%\begin{abstract}
	{\bf The cuprates and iron-based high-temperature superconductors share
		many common features: layered strongly anisotropic crystal structure, strong
		electronic correlations, interplay between different types of
		electronic ordering, the intrinsic spatial inhomogeneity due to doping.
		The understanding of complex interplay between these factors is crucial for
		a directed search of new high-temperature superconductors. Here we show the appearance of inhomogeneous gossamer superconductivity in bulk FeSe compound at ambient pressure and at temperature 5 times higher than its zero-resistance $T_c$. This discovery helps to understand numerous remarkable superconducting properties of FeSe. We also find and prove a general property: if inhomogeneous superconductivity in a anisotropic conductor first appears in the form of isolated superconducting islands, it reduces electric resistivity anisotropically with maximal effect along the least conducting axis. This gives a simple and very general tool to detect inhomogeneous superconductivity in anisotropic compounds, which is critically important to study the onset of high-temperature superconductivity.}\\
%\end{abstract}	

{\bf Introduction}\\
	
A deep understanding of the mechanisms and prerequisites of high-temperature
superconductivity is a fundamental challenge to condensed-matter physics.
In spite of three-decade extensive research, the advance in this field is still
insufficient to reliably predict new high-temperature superconductors.
The maximal superconducting transition temperature $T_{c}$ in the most promising cuprates and iron-based high-temperature
superconductors appears at some non-stoichiometric chemical composition, or
doping\cite{HighTcNatureReview2015,FeBasedHighTcNatureReview2016}.
This inevitably leads to a spatial inhomogeneity of
these compounds because of local variations of doping level.
Hence, high-temperature superconductivity in these compounds, probably, first appears in the form of
small isolated superconducting islands, which become connected and coherent
with decreasing temperature or with changing another driving parameter, i.e.
doping or pressure\cite{KresinReview2006}. Such
inhomogeneous superconductivity with disrupted long-range order is often called as \textit{gossamer
superconductivity}, the term first introduced by Robert Laughlin \cite{Laughlin06}.
Transition to this specific state is supported by the diamagnetic response,
observed in various cuprate superconductors far above $T_{c}$.\cite%
{DiaLa,DiaTl,DiaYBCO1,DiaYBCO} The numerous direct observation of
inhomogeneous electronic structure on a microscopic scale of few nanometers
using STM and other experimental tools was reported in Bi$_{2} $Sr$_{2}$CaCu$%
_{2}$O$_{8+\delta }$,\cite%
{BisccoInhom2001,BisccoInhom2002,BisccoInhom2007,BisccoInhom2009} in HgBa$%
_{2}$CuO$_{4+\delta }$,\cite{HgInhom2015} in Fe-based high-Tc superconductor
Pr-doped CaFe$_{2}$As$_{2}$ ($T_{c}\approx 45$ K),\cite{CaFeAsPRL2014} etc.
The superconductivity in these compounds, probably, develops in two stages:
(i) the preformation of Cooper pairs on isolated islands, which leads to the diamagnetic response and was even proposed to
be an origin of pseudogap in cuprates,\cite{KresinReview2006} and (ii) the
onset of long-range coherence between superconducting islands,
leading to a zero resistance along the whole sample.

Whether the spatial inhomogeneity is a concomitant or assistant feature
of high-temperature superconductivity is still
debated, although various theoretical models propose an enhancement of
superconducting transition temperature due to such inhomogeneity.\cite%
{InhomEnhanceSC,KresinReview2006} It may also play an important role in
thin FeSe films on the interface of SrTiO$_3$, where superconductivity
with high transition temperature $T_{c}\sim 109K$ was reported\cite{Ge15}.
In any case, it is highly desirable to
have a general and simple experimental test if superconductivity first
appears in the form of isolated islands. Such a test would be useful in high-$T_{c}$
superconductors and in many other compounds. The spatial inhomogeneity of
superconductivity may arise not only due to doping, but also due to an
interplay between different types of electronic ordering. For example, the
interplay between spin-density wave at imperfect Fermi-surface nesting and
superconductivity in the organic superconductor (TMTSF)$_{2}$PF$_{6}$ also
leads to inhomogeneous superconductivity and even to the anisotropic
superconductivity onset: superconductivity in this layered compound first
observed along the least conducting axis, perpendicular to conducting planes.%
\cite{PasquierPRB2010,Chaikin} This feature looks odd and counterintuitive,
however, the similar effect was also reported in another organic
superconductor (TMTSF)$_{2}$ClO$_{4}$,\cite{Gerasimenko2014} and in
the cuprate high-Tc superconductor YBa$_{2}$Cu$_{4}$O$_{8}$.\cite{AnisScYBCO}

In this paper we first formulate and prove a general property: if superconductivity in an
anisotropic conductor appears in a gossamer form of disconnected
superconducting islands, these islands reduce electric resistivity
anisotropically, i.e. their influence is first seen in electron transport
along least conducting axis, perpendicular to conducting planes. This
property may show up as an anisotropic superconductivity onset: the electric
resistivity drops considerably along least conducting axis and remain almost
unchanged along conducting planes. Below we substantiate and describe this
property theoretically, and then demonstrate it experimentally on a FeSe
superconductor. Using this property we show that superconductivity in bulk FeSe
at ambient pressure first appears in the form of isolated islands at temperature
$T_{c}^{\ast }\approx 35-40K$, which is close to superconducting transition temperature at high pressure\cite{Mizuguchi08,Medvedev09,FeSePressure} and strongly exceeds the zero-resistance superconducting transition temperature $T_{c}=8K$ at ambient
pressure\cite{Hsu08}. This discovery may help to understand numerous remarkable and unusual superconducting properties of this compound, e.g. the high transition temperature $T_{c}\sim 109K$ in thin FeSe films\cite{Ge15}.\\

{\bf Theoretical description of resistivity drop due to rare
	superconducting islands}\\

%{\bf Simple qualitative model of anisotropic superconductivity onset}

In a layered conductor with the anisotropy parameter $\eta \equiv \sigma
_{zz}/\sigma _{xx}\ll 1$ and small superconducting islands of volume ratio $%
\phi \ll 1$ (see Fig. 1) there are two parallel ways for interlayer current $j$
to flow, so that the total interlayer current and conductivity are approximately given by
the sums of two terms: $j_{tot}=j_{1}+j_{2}$ and $\sigma _{zz}^{tot}=\sigma _{zz}^{(1)}+\sigma _{zz}^{(2)}$.
The first, standard way is with almost uniform current density and
direction $\boldsymbol{j}_{1}\left( \boldsymbol{r}\right) $ perpendicular to
the conducting layers. The rare superconducting inclusions then only slightly
increase corresponding interlayer conductivity $\sigma _{zz}^{(1)}$
proportionally to their volume ratio, and $\sigma _{zz}^{(1)}\sim \eta
\sigma _{xx}$. The second way of interlayer current is via superconducting islands. Since these
superconducting islands are rare, the major part of the current path is in
the normal phase. But instead of flowing along the external field $E_{z}$,
the current between the islands can flow along the highly conducting layers
until it comes to another superconducting island which allows next lift in
the interlayer direction. Then there is no local current density along the $z$%
-axis in the normal phase, and the interlayer conductivity $\sigma _{zz}^{(2)}$ does
not acquire the small anisotropy factor $\eta $. However, its path along conducting
layers between rare superconducting islands is long and inversely
proportional to the volume ratio of superconducting phase $\phi $, so that $%
\sigma _{zz}^{(2)}\sim \phi \sigma _{xx}$. Depending on the ratio $\eta
/\phi $ the first or second way makes the main contribution to the
interlayer conductivity $\sigma _{zz}^{tot}$\ in such a heterogeneous media.

%{\bf Quantitative model with spherical superconducting islands in
%	anisotropic conductor}

In the limit of rare superconducting islands, when their volume fraction $%
\phi \ll 1$, one can apply the Maxwell's approximation (see Sec. 18.1.1 of
Ref. \cite{Torquato}), first proposed by Maxwell in 1873 for isotropic 3D
case. Then the isotropic 3D media of conductivity $\sigma _{1}$ with
spherical inclusions (granules) of conductivity $\sigma _{2}$ with small
volume fraction $\phi \ll 1$ is equivalent to the uniform media with
effective conductivity $\sigma _{e}$ determined by the linear equation
(see Sec. A of Supplementary information for details)
\begin{equation}
	\frac{\sigma _{e}-\sigma _{1}}{\sigma _{e}+2\sigma _{1}}=\phi \frac{\sigma
		_{2}-\sigma _{1}}{\sigma _{2}+2\sigma _{1}},  \label{Maxwell1}
\end{equation}%
which gives%
\begin{equation}
	\frac{\sigma _{e}}{\sigma _{1}}=1+\frac{3\phi \left( \sigma _{2}-\sigma
		_{1}\right) }{\sigma _{2}\left( 1-\phi \right) +\sigma _{1}\left( 2+\phi
		\right) }.  \label{M1}
\end{equation}

	\begin{figure}[tbp]
	\includegraphics[width=0.7\textwidth]{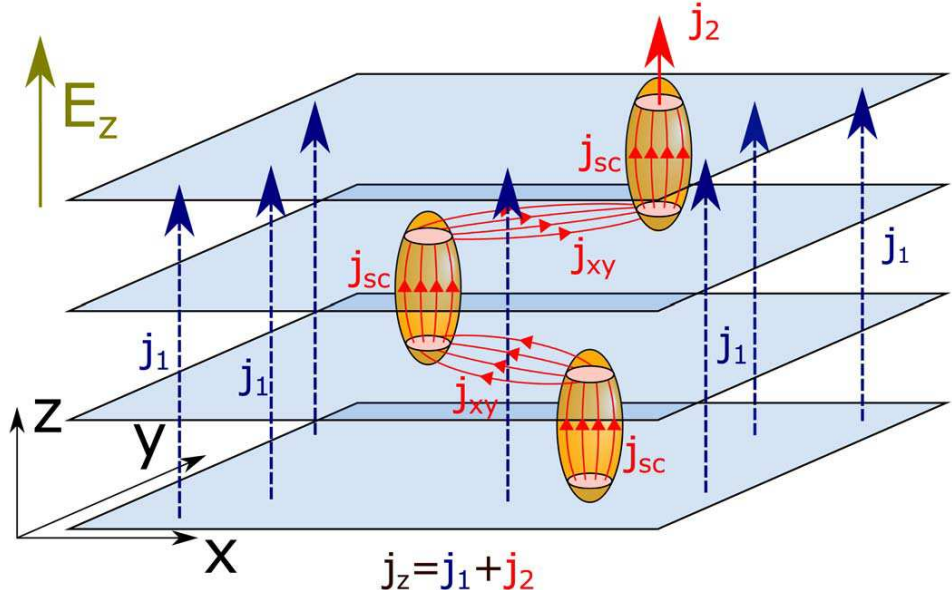}
		\caption{{\bf Illustration of two ways of interlayer current in a heterogeneous media with superconducting inclusions.} The first current path $j_{1}$, shown by blue arrows, is perpendicular to the conducting layers. The second diffusive path $j_{2}$, shown by red arrows, goes via superconducting islands and contains long intralayer tracks. The total interlayer current $j_{z}$ is approximately a sum of these two contribution: $j_{z}\approx j_{1}+j_{2}$. The yellow ellipsoids illustrate superconducting islands.}
		\label{FigScheme}
	\end{figure}

	The problem of conductivity in anisotropic media can be mapped to the
problem of isotropic media with anisotropic coordinate dilations
(see Sec. B of Supplementary information for details). Thus, the
current flow in the media with the easy-plane anisotropy, i.e. where $\sigma
_{zz}\ll \sigma _{xx}=\sigma _{yy}$, is similar to the current flow in
(mapped) isotropic media with $\sigma _{zz}^{\ast }=\sigma _{xx}^{\ast
}=\sigma _{yy}^{\ast }=\sigma _{xx}$ subjected to uniaxial dilation along
the $z$-axis: $z_{\ast }=z/\sqrt{\eta }$, where $\eta =\sigma _{zz}/\sigma
_{xx}$. Then the spherical inclusions inside anisotropic media transform to
elongated ellipsoids with axis ratio $a_{z}/a_{x}=1/\sqrt{\eta }\gg 1$ and
eccentricity $\chi =\sqrt{1-\eta }\rightarrow 1$, which are similar to finite filaments along $c$-axis.
The generalization of Eq. (\ref{Maxwell1}) for the mapped media is\cite{Torquato}
\begin{equation}
	\left( 1-\phi \right) \left( \boldsymbol{\sigma }_{e}^{\ast }-\sigma _{1}%
	\boldsymbol{I}\right) +\frac{\phi \left( \boldsymbol{\sigma }_{e}^{\ast
		}-\sigma _{2}\boldsymbol{I}\right) }{\boldsymbol{I+A}\left( \sigma
		_{2}-\sigma _{1}\right) /\sigma _{1}}=0,  \label{M2}
\end{equation}%
where $\boldsymbol{I}$ is a unitary 3x3 matrix, and the diagonal matrix $%
\boldsymbol{A}$ for prolate spheroidal ($a_{x}=a_{y}$) inclusions is%
\begin{equation}
	\boldsymbol{A=}\left(
	\begin{array}{ccc}
		Q & 0 & 0 \\
		0 & Q & 0 \\
		0 & 0 & 1-2Q%
	\end{array}%
	\right) ,  \label{A}
\end{equation}%
where%
\begin{equation}
	2Q=1+\frac{1}{1/\eta -1}\left[ 1-\frac{1}{2\chi }\ln \left( \frac{1+\chi }{%
		1-\chi }\right) \right] .  \label{Q}
\end{equation}%
For isotropic case $Q=1/3$, the matrix $\boldsymbol{A=I/3}$, and Eq. (\ref%
{M2}) simplifies to Eq. (\ref{Maxwell1}). For strong anisotropy $\eta
=\sigma _{zz}/\sigma _{xx}\ll 1$, the eccentricity $\chi \approx 1-\eta /2$
is close to unity, and
\begin{equation}
	Q\approx 1/2+\eta \left[ 1+\ln \left( \eta /4\right) /2\right] /2.
	\label{Q1}
\end{equation}

Substituting Eq. (\ref{A}) to Eq. (\ref{M2}) gives the linear matrix
equation on $\boldsymbol{\sigma }_{e}^{\ast }$:
\begin{equation}
	\left( 1-\phi \right) \left( \boldsymbol{\sigma }_{e}^{\ast }-\sigma _{1}%
	\boldsymbol{I}\right) \left( \boldsymbol{I}\sigma _{1}\boldsymbol{+A}\left(
	\sigma _{2}-\sigma _{1}\right) \right) +\phi \left( \boldsymbol{\sigma }%
	_{e}^{\ast }-\sigma _{2}\boldsymbol{I}\right) \sigma _{1}=0.  \label{MEq}
\end{equation}%
The solution of this equation is the diagonal matrix $\boldsymbol{\sigma }%
_{e}^{\ast }$. Its three diagonal elements at $\sigma _{2}/\sigma
_{1}\rightarrow \infty $ simplify to
\begin{equation}
	\frac{\sigma _{xx}^{\ast }}{\sigma _{1}}\rightarrow \frac{Q\left( 1-\phi
		\right) +\phi }{Q\left( 1-\phi \right) }=1+\frac{\phi }{Q\left( 1-\phi
		\right) },  \label{sx}
\end{equation}%
$\sigma _{yy}^{\ast }=\sigma _{xx}^{\ast }$, and
\begin{equation}
	\frac{\sigma _{zz}^{\ast }}{\sigma _{1}}\rightarrow \frac{2Q\left( 1-\phi
		\right) -1}{\left( 2Q-1\right) \left( 1-\phi \right) }=\frac{1}{1-\phi }+%
	\frac{2Q\phi }{\left( 1-2Q\right) \left( 1-\phi \right) }.  \label{sz}
\end{equation}%
For strongly anisotropic compounds with $\eta \ll 1$, substituting Eq. (\ref%
{Q1}) and making reverse mapping $z=\sqrt{\eta }z_{\ast }$ and $\sigma
_{zz}=\eta \sigma _{zz}^{\ast }$ to initial problem, from Eqs. (\ref{sx})
and (\ref{sz}) we finally obtain
\begin{equation}
	\sigma _{xx}\approx \sigma _{1}\left( 1+2\phi \right) ,  \label{sxf}
\end{equation}%
and%
\begin{equation}
	\sigma _{zz}\approx \sigma _{1}\left( \frac{\eta }{1-\phi }+\frac{\phi }{\ln
		\left( 2/\sqrt{\eta }\right) -1}\right) .  \label{szf}
\end{equation}%
The expression (\ref{szf}) for interlayer conductivity $\sigma _{zz}$
consists of two parts. The first (regular) part at $\phi \ll 1$ only
slightly increases, similarly to $\sigma _{xx}$. This part corresponds to
the usual interlayer transport with local current density $j_{1}$ almost
perpendicular to conducting layers, so that it contains the small anisotropy
factor $\sigma _{zz}/\sigma _{xx}\approx \eta $. The second (irregular) part
of $\sigma _{zz}$ in Eq. (\ref{szf}) corresponds to the strongly nonuniform
current density $j_{2}$: the current flows via superconducting islands along
$z$-axis and between these superconducting islands along conducting $\left(
x,y\right) $ planes. This term does not have small anisotropy factor $\eta $%
, but contains another small factor $\phi $, the volume fraction of
superconducting islands. Hence, at $\phi >\eta $ the resulting conductivity
anisotropy due to spherical superconducting islands reduces from $\sigma
_{xx}/\sigma _{zz}=1/\eta \gg 1$ to $\sigma _{xx}/\sigma _{zz}\approx 1/\phi
$.\\

{\bf Methods}\\

\begin{figure}[thb]
	\includegraphics[width=0.9\textwidth]{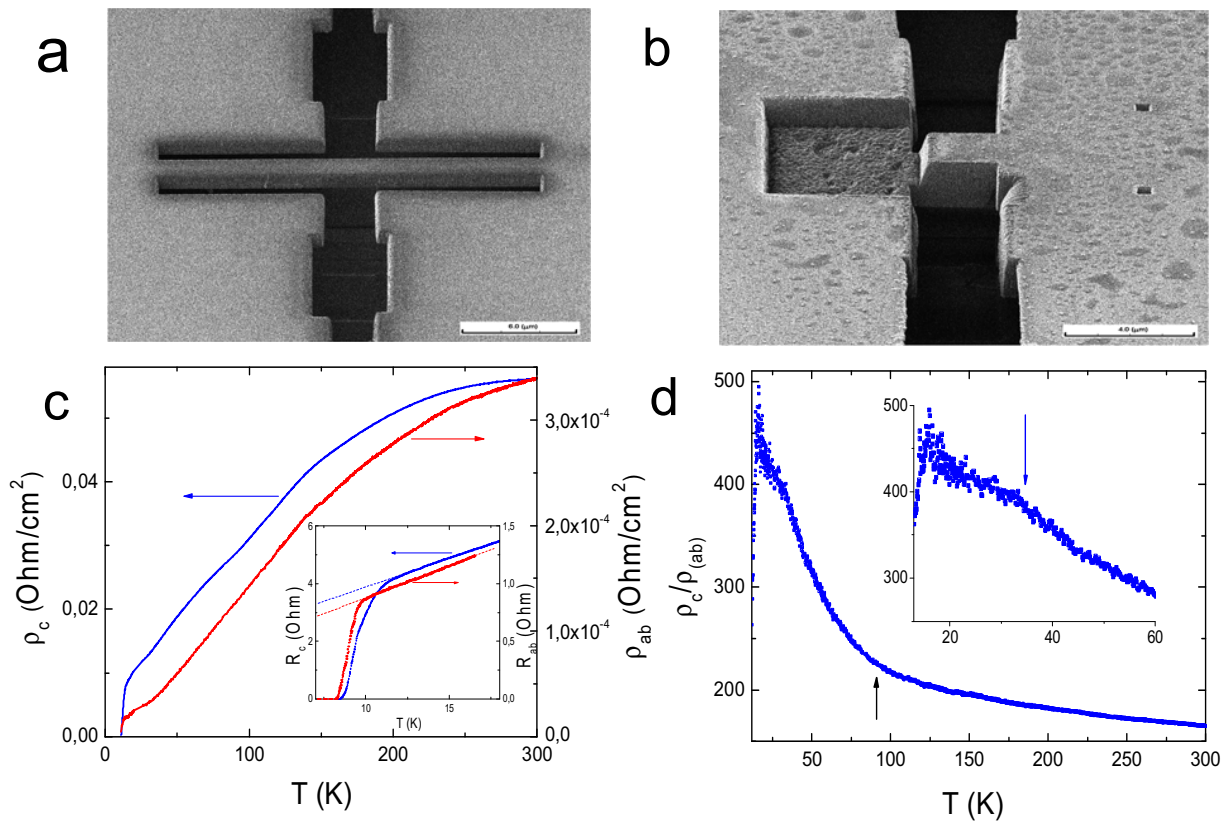}
		\caption{{\bf Intra- and interlayer conductivity measurements.} {\bf a,} The SIM image of FeSe in-plane microbridge;  {\bf b,} The SIM image of FeSe microbridge (overlap structure) oriented along $c$-axis; {\bf c,} temperature dependencies of resistivity: red curve -- structure A and blue curve -- structure B. Inset shows superconducting transition for both types of structures; {\bf d,} Anisotropy of conductivity, $\protect\rho _{c}/\protect\rho _{ab}$ as a function of temperature.}
		\label{F1}
	\end{figure}

For present experiments we have chosen good quality plate-like single crystals (flakes) of FeSe$_{1-\delta }$ superconductor,
grown in evacuated quartz ampoules using AlCl3/KCl flux technique in
permanent temperature gradient, as described in \cite{Chareev2013}. Thin
single-crystal samples with a thickness typically 2-4 $\mu $m were prepared
by micromechanical exfoliation of relatively thick crystals. The structures
of two types have been fabricated by the focused ion beam (FIB) technique
described in \cite{scenm} from selected samples (see Fig. \ref{F1} (a) and
(b)). Structure of the first type, called below as A-type and shown in Fig.\ref{F1}a,
is an in-plane bridge of length 20$\mu$m, width 2 $\mu$m, and thickness
equal to single crystal thickness. This bridge is used to measure the intralayer resistance $\rho_{ab}$. Structure of the second type, shown in Fig.\ref{F1}b and called below as B-type, is a bridge oriented transverse to the layers, along the $c$-axis, with a typical sizes $L_a\times L_b\times L_c=$2$\mu$m$\times$2$\mu$m$\times$0.2 $\mu$m. This bridge is used to measure the interlayer resistance $\rho_c$. The electrical contacts to the crystal have been prepared by the laser evaporation of gold films before the
processing by FIB. The measurements of electrical resistance and of current-voltage (IV)
characteristics have been done in the conventional 4-probe configuration. The
temperature dependence of magnetic susceptibility of FeSe single crystal was
obtained by AC Measurement option of Physical Property Measurement System --
9T Quantum Design. The plate was oriented perpendicular to
external magnetic field $H=10$ Oe applied at frequency 10 kHz. The
demagnetizing factor $N\sim0.5$ was supposed to obtain the full Meissner effect $%
4\pi\chi=-1$ for a finite size plate in accordance with classical formula \cite%
{Goldfarb91}.\\
	
{\bf Results and discussion}\\

Fig.\ref{F1} (c) shows the temperature dependence of resistivity in the structures of both types.
The well-defined geometry of these structures allowed us to determine the conductivity anisotropy ratio $\rho_c/\rho_{ab}$ and
its temperature evolution, shown in Fig.\ref{F1}d. At room temperature $%
\rho_c/\rho_{ab}\approx 160-180$ and increases monotonically with temperature,
reaching $\approx 500$ at $T=12$K. Note, that this increase of anisotropy
goes in two stages. First, in the temperature range 300-90 K, the rate of this increase is about $%
0.25-0.30$ K$^{-1}$. Then, below 90 K, this rate increases by one order of magnitude, achieving 2.5-3.0 K$^{-1}$.
Such a behavior reflects the strong decreasing of interlayer conductivity at temperature below the
structural transition.\cite{McQueen09,Terashima15} Additionally, in Fig. Fig.\ref{F1}d one may see a small kink at $T\approx 35K$. This feature is discussed below in detail. We suggest that it comes due to the appearance of inhomogeneous superconductivity in the form of isolated microscopic islands.
	\begin{figure}[thb]
		\includegraphics[width=0.9\textwidth]{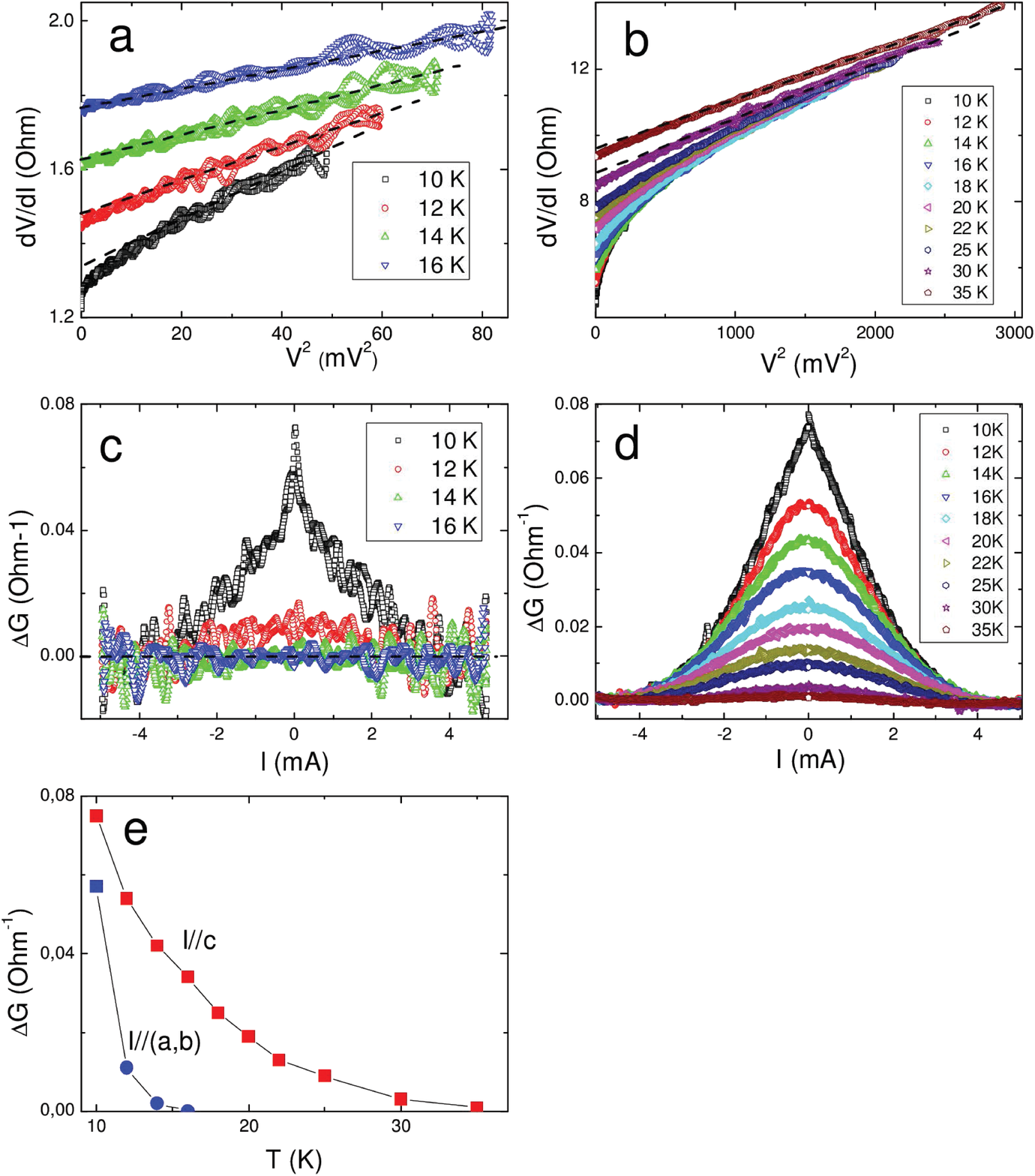}
		\caption{{\bf Excess conductivity for two perpendicular current directions.} {\bf a,} Differential resistance $dV/dI$ as a function of the
			square of voltage, $V^2$, at different temperatures above $T_c$ for A-type
			structure, i.e. for intralayer transport. {\bf b,} The same as (a) but for B-type structure, i.e. for interlayer transport. {\bf c,}
Excess differential conductivity as a function of current for A-type structure
			measured at different temperatures above $T_c$. {\bf d,} The same as in (c) but
			for B-type structure. {\bf e,} Temperature dependence of excess conductivity:
			red square symbols for the B-type structure, and blue circle symbols for the A-type structure.
			%f) Temperature dependencies of the volume fraction of superconducting phase, $%
			%\protect\phi$, calculated from equation \protect\ref{szf}.
		}
		\label{F2}
	\end{figure}
Taking into account the layered crystal structure of FeSe, in B-type structures one may expect to observe
some effects of weak superconductivity, namely,
intrinsic Josephson effect, similar to that in layered cuprate
high-T$_c$ superconductors \cite{Kleiner94}. Surprisingly, in our structures
we have observed just opposite picture: superconductivity is stronger in the direction perpendicular to conducting layers
as compared to intralayer superconductivity.
Inset in Fig.\ref{F1}c demonstrates superconducting transition for both
types of structures. As can be seen, superconducting temperature in B-type
structure is higher than in A-type structure. Such a behavior was observed
for all studied samples. Moreover, the critical current density in the B-type
structures is also larger by an order of magnitude.

Most interesting result was obtained during the study of the current-voltage (IV)
characteristics of the bridge structures at temperature above $T_c$. Typical linear $R(T)$ behavior in a normal metallic state corresponds to the quadratic dependence of differential resistance on voltage because of the small Joule heating. In our case one may expect small deviations from this square dependence caused by superconducting fluctuations which appear in IV curves as excess conductivity at temperature close to $T_c$. Fig.\ref{F2} (a) and (b) illustrates $dV/dI$ as a function of $V^2$ at different temperatures above $T_c$ for the A-type and B-type structures respectively. Figs. \ref{F2}c and \ref{F2}d show corresponding excess conductivity as a function of current, which was obtained by the extracting of normal state quadratic background from the experimental IV curves. One sees that the intralayer electronic transport in A-type structure demonstrates conventional for superconductors behavior: excess conductivity and, correspondingly, the superconducting fluctuations disappear rapidly above bulk $T_c$ and they are absent completely above $T=13$ K. This result correlates well with $R(T)$ behavior for this type structures (inset in Fig. \ref{F1}c).

Quite different behaviour is observed in the interlayer electronic transport, i.e. in the B-type structure. As can be seen from Fig. \ref{F2}(b,c), the excess conductivity is much more pronounced and, more importantly, observed up to $T\approx 35$ K. Note, that $R(T)$ for this type of junction is strongly linear at least in the temperature range 14-25 K (see inset in Fig.\ref{F1}c). It means, that simple fluctuation effects cannot cause the observed excess conductivity. The difference between intralayer and interlayer conductivity is clearly seen in Fig.\ref{F2}e, where we plot the temperature dependence of excess conductivity for both types of structures.

We see only one explanation of the observed effect: the formation of small
superconducting islands with $T_c^*\approx35-40$ K. Then at $%
T=T_c^*$ the corresponding decrease of resistance $R(T)$ should be anisotropic
according to the above theoretical model.

Our results are in agreement with the recent work \cite{Naidyuk} where a rise in $%
T_{c}$ more than twice was observed in point-contacts between FeSe
single crystal and Cu. Actually, authors of this work observed some excess
conductivity at temperature well above $T_{c}$. The point-contact was formed
between cooper wire and the plane of FeSe crystal. It is well known that the
point-contact itself is directional with respect to the electric-field
configuration \cite{sinch2003}, and one may expect that the main contribution
to the point-contact resistance comes from injection along the point-contact
orientation, making point contact configuration close to our structures.
Then the authors of Ref. \cite{Naidyuk} probed the electronic transport mainly along the $c$%
-axis and, therefore, observed similar effect from the filamentary gossamer
superconductivity.

%{\bf Indications of high-T superconducting transition. Diamagnetic
%	response and resistivity peculiarity.}

	\begin{figure}[thb]
		\includegraphics[width=1.0\textwidth]{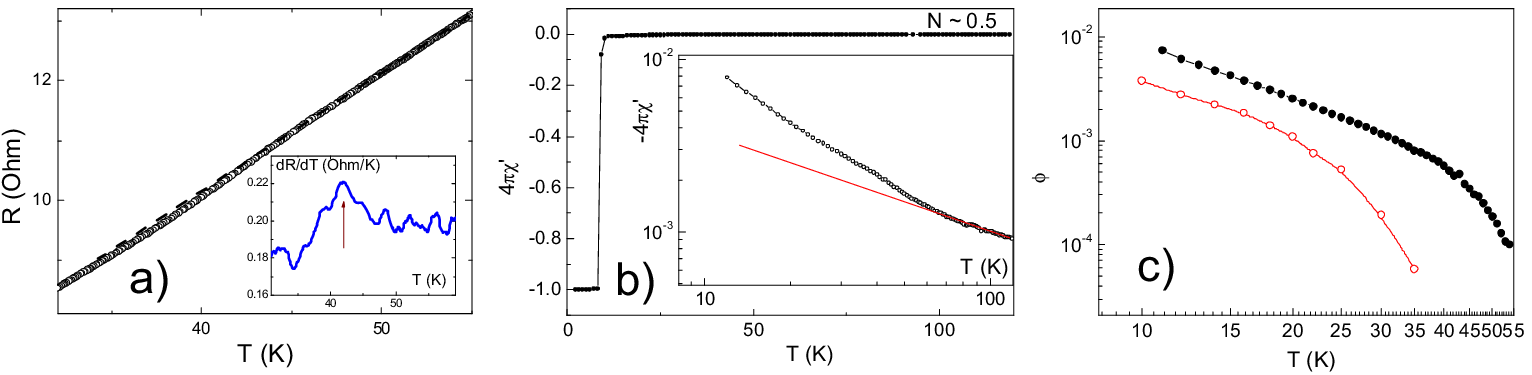}
		\caption{{\bf Temperature dependence of the volume fraction of superconducting islands.} {\bf a,} Temperature dependence of interlayer resistance in the temperature range 30-60 K demonstrating small decrease of resistance
			at $T\approx 45$ K. Inset shows the derivative of this curve. {\bf b,} The temperature
			dependence of real part of magnetic susceptibility of FeSe single crystal.
			Main panel contains initial $4\protect\pi \protect\chi $ curve obtained for
			demagnetizing factor $N \sim 0.5$, and the inset represents the same curve in double
			logarithmic scale to highlight the negative deviation at high temperatures.
			The line is a guide for an eye. {\bf c,} The temperature dependence of
			superconducting volume fraction $\phi $ obtained from magnetic (closed circles) and
			from transport (open circles) measurements. In the latter case, the volume fraction $\protect\phi $
			of superconducting phase was calculated from equation (\ref{szf}).}
		\label{F4}
	\end{figure}
	
We observe small peculiarities already in the $R(T)$ dependencies
which may indicate the appearance of superconducting islands.
A very small but visible decrease of interlayer resistance can be noticed
at $T\approx 42-45$ K as shown in Fig. \ref{F4}(a). This effect is more
pronounced in the derivative curve, $dR/dT(T)$, shown in the inset to Fig. \ref{F4}(a).
Note, that this feature is completely absent in the intralayer
resistance. As one can see in the inset in Fig.\ref{F1}(d), at nearly the same
temperature some decrease of anisotropy is also observed.

One of the best ways to detect the existence of small volume fraction of
superconductivity is the measurement of magnetic susceptibility. In the present
work we also measured the magnetic properties of studied crystals. The temperature dependence of the real part of magnetic susceptibility $4\pi\chi$ shown in Fig. \ref{F4} (b) demonstrates a negative deviation in the whole temperature range. At high temperatures it decreases almost linearly then bends down at approximately 50 K which can be seen more
clearly in double logarithmic scale shown in the Inset to Fig. \ref{F4} (b),
and finally drops to absolute diamagnetic value $4\pi\chi=-1$ below
superconducting phase transition $T_C=9$ K. The rough estimation of
superconducting phase portion $\phi$ can be done by subtraction of linear
function from $4\pi\chi(T)$ dependence assuming 100\% of superconducting
phase at low temperatures $T\ll T_C$. The temperature dependencies of
superconducting phase percentage obtained from magnetic (closed circles) and
transport (open circles) measurements are shown in Fig. \ref{F4} (c). In the
latter case, the volume fraction $\phi$ of superconducting phase was
calculated from Eq. \ref{szf} using the experimental data on conductivity
anisotropy, shown in Fig. \ref{F1}d. Fig. \ref{F4}(c) shows that $\phi$ is very
small and decreases monotonically with increasing temperature. It is
distinguishable below 50 K and amounts $10^{-4}$ at this temperature. At
lower temperatures it is somewhat higher and comprises $10^{-2}$ above $T_c$.
 The shape of $\phi (T)$ dependence obtained from magnetic measurements is
similar to that from transport measurements (see Fig. \ref{F4}c).\\

{\bf Conclusions}\\

In this paper we report the discovery of inhomogeneous superconductivity in bulk FeSe$_{1-\delta }$ at
ambient pressure and at temperature $T_{c}^{\ast }\sim 35K$, which is 5
times higher than its zero-resistance superconducting transition temperature $T_{c}\approx 8K$ known before\cite{Hsu08}. This superconductivity appears in the form of microscopic isolated superconducting islands and does not lead to zero electric resistance, but reveals itself in anisotropic resistivity drop and in magnetic susceptibility. Therefore, we called it gossamer inhomogeneous
superconductivity. This discovery provides a clue to understand
the nature of unusual superconducting properties of FeSe. For example, the
pressure-temperature phase diagram of FeSe contains interplay between
different electronic ordering and superconductivity with $T_{c}\sim 35K$ at pressure 6-8GPa,\cite%
{Mizuguchi08,Medvedev09,FeSePressure} which is the same as the transition temperature $T_{c}^{\ast }$ of gossamer inhomogeneous
superconductivity reported above.
This coincidence is not occasional but suggests that the external pressure, by damping other types of electronic ordering, increases the volume fraction of superconducting regions, so that they become connected at pressure 6-8GPa. However, even at ambient pressure in FeSe there are microscopic superconducting islands without long-range coherence between these islands and, hence, without zero resistance.

Complementary, we proposed and described a general property: if inhomogeneous
superconductivity in a anisotropic conductor first appears in the form of
isolated superconducting islands, it reduces electric resistivity
anisotropically with maximal effect along the least conducting axis. This property
provides a simple and very general tool to detect inhomogeneous
superconductivity in various anisotropic compounds. Namely, this
method is applicable to almost all high-temperature superconductors, which
have layered anisotropic crystal structure. The above study of FeSe is a nice illustration of this
general tool. The anisotropic resistivity drop during the superconductivity onset was also
reported in other layered superconductors, e.g. in organic superconductors
(TMTSF)$_{2}$PF$_{6}$ \cite{PasquierPRB2010,Chaikin} and (TMTSF)$_{2}$ClO$%
_{4}$ \cite{Gerasimenko2014}, and in even in the cuprate high-Tc
superconductor YBa$_{2}$Cu$_{4}$O$_{8}$ (see Fig. 2 in Ref. \cite{AnisScYBCO}%
). Our model and Eq. (\ref{szf}) explains these observations and suggests
inhomogeneous superconductivity onset also in these compounds. We believe that
similar experimental test for inhomogeneous superconductivity can be
performed in many other anisotropic superconductors, which will help to understand the mechanisms of high-temperature superconductivity.\\
	
	{\bf Acknowledgements}\\
	
	P.G. thanks Dmitrii Lyubshin for useful discussion. This work was supported
	in part by the Ministry of Education and Science of the Russian Federation
	in the framework of Increase Competitiveness Program of NUST
	MISiS (\# K2-2015-075, \# K4-2015-020 and \# K2-2016-003) and by Act 211 Government of the Russian Federation,
	contract \# 02.A03.21.0006. Theoretical part was supported by RSCF \# 16-42-01100.\\
	
	{\bf Author contributions}\\
	
	A.A.S., A.O. and A.F. prepared the samples and performed the electronic transport measurements. P.G. proposed the theoretical description of the observed transport properties. O.V., A.S., D.Ch. and A.V. grew the crystals and performed the magnetic susceptibility measurements. All authors contributed to discussions.

	\newpage

\end{document}

{\bf \textit{Supplementary Material} to {\large ''Gossamer bulk high-temperature superconductivity in FeSe''}}\\
	
	{\bf A. Derivation of the effective conductivity in the Maxwell's
		approximation for heterogeneous media with spherical granules}\\
	
	The effective conductivity $\sigma _{e}$ of heterogeneous media, where small granules of conductivity $\sigma_2$ are embedded  in a media of conductivity $\sigma_1$,
can be easily derived in the effective-medium Maxwell's approximation (see \S 18.1.1 of
Ref. \cite{Torquato}). In this approximation a heterogeneous media is replaced by a uniform media of
	effective conductivity $\sigma _{e}$ such that their electric potentials at large distance are the same.
In the Maxwell's approximation the interaction between rare small granules is neglected, and
	in the external uniform electric field $\boldsymbol{E}_{0}$ each small
	granule of radius $R_{i}$ polarizes and acquires an additional electric
	dipole moment $\boldsymbol{d}_{i}$ proportional to its volume and to the
	strength of the field $E_{0}$:
	\begin{equation}
		\boldsymbol{d}_{i}\boldsymbol{=}\mathbb{\beta }_{12}\boldsymbol{E}%
		_{0}R_{i}^{3},  \label{dd}
	\end{equation}%
	where (see \S 17.1.1 of Ref. \cite{Torquato})
	\begin{equation}
		\beta _{12}=\left( \sigma _{2}-\sigma _{1}\right) /\left( \sigma
		_{2}+2\sigma _{1}\right)  \label{beta}
	\end{equation}%
	is the \textquotedblright polarizability\textquotedblright\ of a sphere.
	Each such dipole moments changes the electric potential outside the granule
	by
	\begin{equation}
		\Delta \varphi _{i}=\boldsymbol{d}_{i}\boldsymbol{r}/r^{3},  \label{dfi}
	\end{equation}%
	so that the total change of electric potential far away from the sphere $%
	R_{0}$, i.e. at $r\gg R_{0}$ is given by the sum of all granules inside
	inhomogeneous sphere of radius $R_{0}$:%
	\begin{equation}
		\Delta \varphi _{t}=\sum_{i}\frac{\boldsymbol{d}_{i}\left( \boldsymbol{r-r}%
			_{i}\right) }{\left\vert \boldsymbol{r-r}_{i}\right\vert ^{3}}\approx \frac{%
			\sigma _{2}-\sigma _{1}}{\sigma _{2}+2\sigma _{1}}\frac{\phi R_{0}^{3}}{r^{2}%
		}E_{0}.  \label{dft}
	\end{equation}%
	On the other hand, a single isotropic sphere of the radius $R_{0}$ and
	conductivity $\sigma _{e}$ inside a media of conductivity $\sigma _{1}$ in
	a uniform field $E_{0}$ creates an additional potential%
	\begin{equation}
		\Delta \varphi _{t}=\frac{\sigma _{e}-\sigma _{1}}{\sigma _{e}+2\sigma _{1}}%
		\frac{R_{0}^{3}}{r^{2}}E_{0}.  \label{dft2}
	\end{equation}%
	Comparing Eqs. (\ref{dft}) and (\ref{dft2}) gives
 \begin{equation}
	\frac{\sigma _{e}-\sigma _{1}}{\sigma _{e}+2\sigma _{1}}=\phi \frac{\sigma
		_{2}-\sigma _{1}}{\sigma _{2}+2\sigma _{1}},  \label{Maxwell0}
\end{equation}%
which coincides with Eq. (1) of the main text.\\

	{\bf B. Mapping of conductivity problem in anisotropic media to isotropic}\\
	
	The electrostatic equation of continuity in Cartesian coordinates can be
	written as%
	\begin{equation}
		-\boldsymbol{\nabla j}=\sigma _{xx}\frac{\partial ^{2}V}{\partial x^{2}}%
		+\sigma _{yy}\frac{\partial ^{2}V}{\partial y^{2}}+\sigma _{zz}\frac{%
			\partial ^{2}V}{\partial z^{2}}=0.
	\end{equation}%
	By the change of coordinates
	\begin{equation}
		x_{\ast }=x,~y_{\ast }=\sqrt{\sigma _{yy}/\sigma _{xx}}y,~z_{\ast }=\sqrt{%
			\sigma _{zz}/\sigma _{xx}}z  \label{CT}
	\end{equation}%
	and by the simultaneous conductivity change $\sigma _{zz}^{\ast }=\sigma
	_{yy}^{\ast }=\sigma _{xx}^{\ast }=\sigma _{xx}$ it transforms to the
	electrostatic continuity equation for isotropic media:
	\begin{equation}
		-\boldsymbol{\nabla j}=\sigma _{xx}\left( \frac{\partial ^{2}V}{\partial
			x_{\ast }^{2}}+\frac{\partial ^{2}V}{\partial y_{\ast }^{2}}+\frac{\partial
			^{2}V}{\partial z_{\ast }^{2}}\right) =0.
	\end{equation}%
	Hence, the initial problem of conductivity in anisotropic media with some
	boundary conditions can be mapped to the conductivity problem in isotropic
	media with new boundary conditions, obtained from initial boundary
	conditions by the anisotropic dilatation in Eq. (\ref{CT}).